\input{aipcheck}

\documentclass[
    ,final            
  ]
  {aipproc}

\layoutstyle{8x11single}

\begin{document}

\title{Stability of $P_{11}$ Resonance Extracted from $\pi N$ Data}


\author{S. X. Nakamura}{
  address={Excited Baryon Analysis Center (EBAC),
Jefferson Laboratory,
Newport News, Virginia 23606, USA}
}

\begin{abstract}
We address a question about
how much resonance poles and residues extracted from data depend on a model
used for the extraction, and on the precision of data. 
We focus on the $P_{11}$ $\pi N$ scattering and use
the dynamical coupled-channel (DCC) model developed in
Excited Baryon Analysis Center (EBAC) at JLab. 
We examine the model-dependence of the poles by varying parameters
largely within the EBAC-DCC model.
We find that two poles associated with the Roper resonance are fairly
stable against the variation. 
We also study 
the stability of the Roper poles against different
analytic structure of the $P_{11}$ amplitude below $\pi N$ threshold
by using a bare nucleon model.
We again find a good stability of the Roper poles. 
\end{abstract}

\maketitle


\section{Introduction}

Extraction of $N^*$ information, such as pole positions and vertex form
factors, is an important task in hadron physics.
This is because they are necessary information to address a question
whether we can understand baryon resonances within QCD.
In order to extract the $N^*$ information, first, one needs to construct
a reaction model through a comprehensive analysis of data.
Then, pole positions and vertex form factors are extracted from the
model with the use of the analytic continuation.
Therefore, the $N^*$ information extracted in this manner is inevitably
model-dependent. 
There are several different approaches to extract the $N^*$ information.
Although almost all
4-stars nucleon resonances  listed by Particle Data Group (PDG)
are found
in all approaches, existence of some $N^*$ states, in particular
those in the higher mass region, is controversial.
Thus, commonly asked questions are how much model-dependent the
extracted resonance parameters are, and how precise data have to be for
a stable resonance extraction.
These are the questions we address at
Excited Baryon Analysis Center (EBAC) at JLab\cite{hnls10},
within a dynamical coupled-channels model (EBAC-DCC)~\cite{msl07}. 
We focus on the $\pi N$ $P_{11}$ partial wave and 
the stability of its pole positions, particularly those corresponding to the
Roper resonance.
In the region near Roper $N(1440)$, two poles
close to the $\pi\Delta$ threshold were found in 
our recent extraction~\cite{sjklms10} from the JLMS model~\cite{jlms07} 
(JLMS is one of EBAC-DCC model),
while only one pole in the
similar energy region was reported in some other analyses.
We examine the stability of this two-pole structure against the
following variation, keeping a good reproduction of 
SAID single-energy (SAID-SES) solution~\cite{said-1} unless otherwise stated.
\begin{itemize}
 \item Large variation of the parameters of the meson-baryon
        and bare $N^*$ parameters of the EBAC-DCC model.
 \item Inclusion of a bare nucleon state:
The analytic structure
of this model is  rather different from the original EBAC-DCC model, in
       particular in the region near the nucleon pole~\cite{jklmss09},  .
 \item Fit to the solution based on the
Carnegie-Mellon University-Berkeley model (CMB)~\cite{cw90} 
which has rather different behavior from SAID-SES for higher $W$.
\end{itemize}

\section{Dynamical coupled-channels models}

Here, we briefly describe dynamical coupled-channels models;
the EBAC-DCC model and the bare nucleon model.

The EBAC-DCC model contains $\pi N$, $\eta N$ and $\pi\pi N$ channels
and the $\pi\pi N$ channel has $\pi\Delta$, $\rho N$ and $\sigma N$
components.
These meson-baryon (MB) channels are connected with each other by meson-baryon
interactions ($v_{MB,M'B'}$), or excited to bare $N^*$ states by vertex
interactions ($\Gamma_{MB\leftrightarrow N^*}$).
With these interactions, 
the partial-wave amplitude for the
$M(\vec{k})+B(-\vec{k}) \to M'(\vec{k}')+B'(-\vec{k}')$
reaction can be written by the following form:
\begin{eqnarray}
T_{MB,M'B'}(k,k',E)  &=&  t_{MB,M'B'}(k,k',E) + t^{R}_{MB,M'B'}(k,k',E),
\label{eq:tmbmb}
\end{eqnarray}
where the first term is obtained by solving the following coupled-channels
Lippmann-Schwinger equation:
\begin{eqnarray}
t_{MB,M^\prime B^\prime}(k,k',E) &=&  v_{MB,M^\prime B^\prime}(k,k')
+ \sum_{M^{\prime\prime}B^{\prime\prime}}
\int_{C_{M^{\prime\prime}B^{\prime\prime}}} \!\!\!\!\!\! q^2 dq
v_{MB,M^{\prime\prime}B^{\prime\prime}}(k,q)
G_{M^{\prime\prime}B^{\prime\prime}}(q,E)
t_{M^{\prime\prime}B^{\prime\prime},M^\prime B^\prime}(q,k',E).
\label{eq:cc-mbmb}
\end{eqnarray}
Here $C_{MB}$ is the integration contour in the complex-$q$ plane used
for the channel $MB$.
The second term of Eq.~(\ref{eq:tmbmb}) is 
associated with the bare $N^*$ states, and given by
\begin{eqnarray}
t^{R}_{MB,M^\prime B^\prime}(k,k',E)&=& \sum_{i,j}
\bar{\Gamma}_{MB \to N^*_i}(k,E) [D(E)]_{i,j}
\bar{\Gamma}_{N^*_j \to M^\prime B^\prime}(k',E),
\label{eq:tmbmb-r}
\end{eqnarray}
where the dressed vertex function 
$\bar{\Gamma}_{N^*_j \to M^\prime B^\prime}(k,E)$ is 
 calculated by convoluting
the bare vertex ${\Gamma}_{N^*_j \to M^\prime B^\prime}(k)$ with
the amplitudes $t_{MB,M^\prime B^\prime}(k,k',E)$.
The inverse of the propagator of dressed $N^*$ states in
Eq.~(\ref{eq:tmbmb-r})
is \begin{equation}
[D^{-1}(E)]_{i,j} = (E - m^0_{N^*_i})\delta_{i,j} - \Sigma_{i,j}(E) ,
\label{eq:nstar-selfe}
\end{equation}
where $m^0_{N^*_i}$  is the bare mass of the $i$-th $N^*$ state,
and the $N^*$ self-energy is defined by
\begin{equation}
\Sigma_{i,j}(E)= \sum_{MB} \int_{C_{MB}}  q^2 dq 
\bar{\Gamma}_{N^*_j \to M B}(q,E) G_{MB}(q,E) {\Gamma}_{MB \to N^*_i}(q,E).
\label{eq:nstar-g}
\end{equation}
Defining  $E_\alpha(k)=[m^2_\alpha + k^2]^{1/2}$ with $m_\alpha$ being
the mass of particle $\alpha$,
the meson-baryon propagators in the above equations are:
$G_{MB}(k,E)=1/[E-E_M(k)-E_B(k) + i\epsilon]$ for the stable
$\pi N$ and $\eta N$ channels, and $G_{MB}(k,E)=1/[E-E_M(k)-E_B(k) -\Sigma_{MB}(k,E)]$
for the unstable $\pi\Delta$, $\rho N$, and $\sigma N$ channels.
The self energy $\Sigma_{MB}(k,E)$ is calculated from a vertex
function defining the decay of the considered unstable particle
in the presence of a spectator $\pi$ or $N$ with momentum $k$.
For example, we have for the $\pi\Delta$ state,
\begin{eqnarray}
\Sigma_{\pi\Delta}(k,E) &=&\frac{m_\Delta}{E_\Delta(k)}
\int_{C_3} q^2 dq \frac{ M_{\pi N}(q)}{[M^2_{\pi N}(q) + k^2]^{1/2}}
\frac{\left|f_{\Delta \to \pi N}(q)\right|^2}
{E-E_\pi(k) -[M^2_{\pi N}(q) + k^2]^{1/2} + i\epsilon},
\label{eq:self-pid}
\end{eqnarray}
where $M_{\pi N}(q) =E_\pi(q)+E_N(q)$ and $f_{\Delta \to \pi N}(q)$
defines the decay of the $\Delta \to \pi N$ in the rest frame
of $\Delta$, $C_3$ is the corresponding integration contour in the
complex-$q$ plane.

To examine further the model dependence of resonance extractions, 
it is useful to also  perform
analysis using models with a bare nucleon, as developed in
Ref.~\cite{afnan}.
Within the formulation of EBAC-DCC model,
such a model can be obtained by
adding a bare nucleon ($N_0$) state with mass $m^0_N$
and $N_0\rightarrow MB $ vertices and
removing the direct $MB \rightarrow N \rightarrow M'B'$
in the meson-baryon
interactions $v_{MB,M'B'}$.
All numerical procedures for this model
are identical to that used for the EBAC-DCC model,
except that the resulting amplitude must satisfy the nucleon pole
condition:
\begin{equation}
t^R_{\pi N,\pi N}(k\to k_{\rm{on}},k\to k_{\rm{on}},E\rightarrow m_N ) 
= -\frac{[F_{\pi NN} (k_{\rm{on}})]^2}{E - m^0_N - \tilde{\Sigma}(m_N)} .
\label{eq:pole-t}
\end{equation}
with 
\begin{eqnarray}
m_N= m^0_N + \tilde{\Sigma}(m_N)
\qquad {\rm and} \qquad
F_{\pi NN}(k_{\rm{on}})= F^{\rm{phys.}}_{\pi NN}(k_{\rm{on}}) \ .
\label{eq:pole-m}
\end{eqnarray}
Here we have used the on-shell momentum defined by
$E=\sqrt{m_N^2+k^2_{\rm{on}}}+\sqrt{m_\pi^2+k^2_{\rm{on}}}$.
Also, $\tilde{\Sigma}(m_N)$ is
the self-energy for the nucleon.
More details for the calculational 
procedure following Afnan and Pearce
is found in Refs.~\cite{hnls10,afnan}.

\section{Results}

Now we show our numerical results to examine the stability of the
$P_{11}$ poles. 
In the following subsections, we present results from various fits by
varying the dynamical content of the EBAC-DCC model,
and by using a model with a bare nucleon.
We show in figures the quality of fits of these models, and 
in Table \ref{tab:p11-tab1} the pole positions from the models
as well as $\chi^2$ per data point ($\chi^2_{pd}$).
We find the poles with the
method of analytic continuation discussed in detail
in Refs.~\cite{sjklms10,ssl09}.
In Table \ref{tab:p11-tab1}, we also present pole positions from 
JLMS\cite{jlms07} and SAID-EDS (energy-dependent)\cite{said-1}.

\begin{table}[h]
\renewcommand{\arraystretch}{1.3}
\tabcolsep=3.0mm
\begin{tabular}{ccccccc}\hline
Model           & $upuupp$   & $upuppp$  & $uuuupp$  & $uuuuup$& $\chi^2_{pd}$   \\ 
\hline
SAID-EDS        & (1359, 81) & (1388, 83) &    ---    &  ---        & 2.94 \\
JLMS            & (1357, 76) & (1364, 105)&    ---    & (1820, 248) & 3.55 \\
2$N^*$-3p       & (1368, 82) & (1375, 110)&    ---    & (1810, 82)  & 3.28 \\
2$N^*$-4p       & (1372, 80) & (1385, 114)& (1636, 67)& (1960, 215) & 3.36 \\
\hline
1$N_0$1$N^*$-3p & (1363, 81) & (1377, 128)&    ---    & (1764, 137) & 2.51 \\\hline
\end{tabular}
\caption{\label{tab:p11-tab1}
The resonance pole positions $M_R$ for $P_{11}$
[listed as ($\rm{Re}M_R$, $-\rm{Im} M_R$) in the unit of MeV] extracted from
 various parameter sets.
The location of the pole is specified by, e.g.,
$(s_{\pi N},s_{\eta N},s_{\pi\pi N},s_{\pi\Delta},s_{\rho N},s_{\sigma N})=(upuupp)$,
where $p$ and $u$ denote the physical and unphysical sheets for a
given reaction channel, respectively. $\chi^2_{pd}$ is $\chi^2$ per data point.}
\end{table}

\subsection{2$N^*$-3p and 2$N^*$-4p fits}
\label{sec3b}

\begin{figure}[t]
\begin{minipage}[t]{75mm}
   \includegraphics[width=75mm]{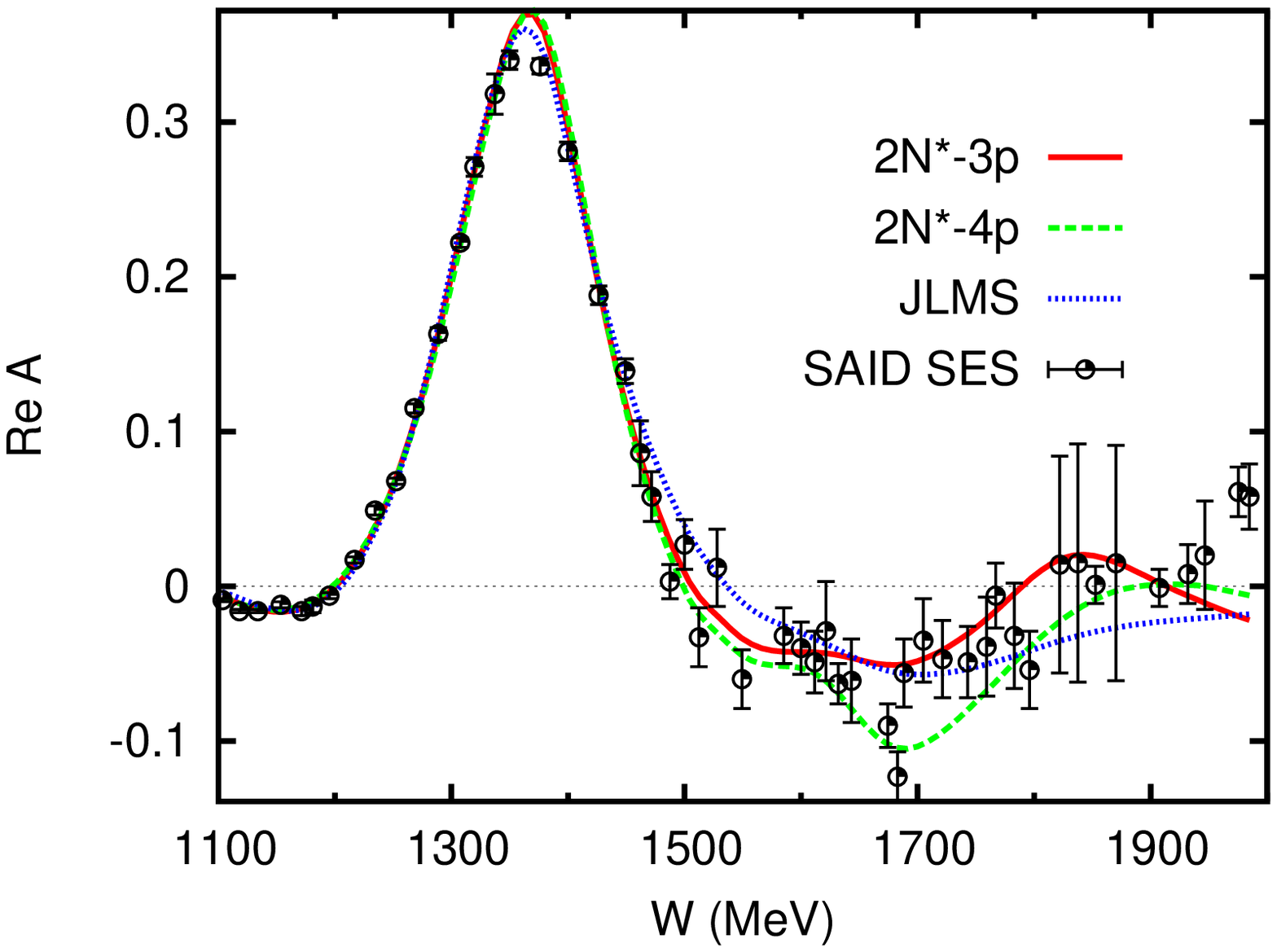}
 \end{minipage}
\begin{minipage}[t]{75mm}
   \includegraphics[width=75mm]{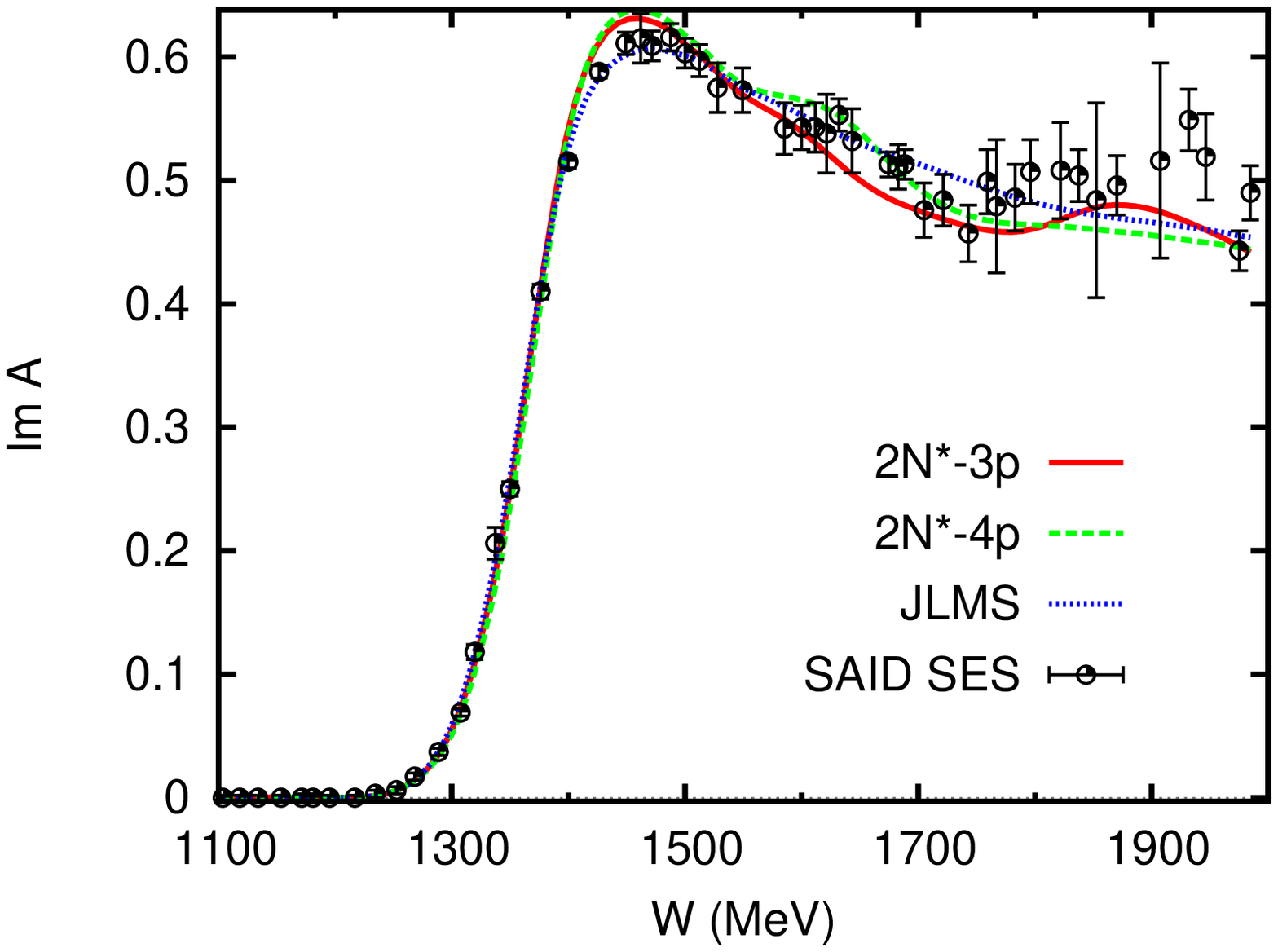}
 \end{minipage}
\caption{\label{fig:p11-fig2}
The real (left) and imaginary (right) parts of the on-shell $P_{11}$
amplitudes as a function of the $\pi N$ invariant mass $W$ (MeV).
$A$ is unitless in the convention of Ref.~\cite{said-1}.
}
\end{figure}

We varied both the parameters for the meson-baryon interactions
($v_{MB,M'B'}$) and parameters associated with bare $N^*$ states 
($m_{N^*}^0$, $\Gamma_{N^*\leftrightarrow MB}$).
The obtained meson-baryon interactions are quite different from those of
JLMS.
We obtained several fits which are different in how the oscillatory
behavior of SAID-SES amplitude for higher $W$ is fitted.
The results from the 2$N^*$-3p (dotted curves) and 2$N^*$-4p (dashed curves)
fits are compared with the JLMS fit (solid curves) 
in Fig.~\ref{fig:p11-fig2}.
The resulting resonance poles
are listed in the 3th and 4th rows of Table~\ref{tab:p11-tab1}.
Here we see again the first two poles near the $\pi\Delta$ threshold from
both fits agree well with the JLMS fit. This seems to further support the
conjecture that these two poles are mainly sensitive to the
data  below $W\sim 1.5$ GeV where the SAID-SES has rather small errors.
However, the 2$N^*$-4p fit has one more pole at 
$M_R= 1630 -i45$ MeV. This
is perhaps related to its oscillating structure near $W\sim 1.6$ GeV
(dashed curves), as shown in the Figs.~\ref{fig:p11-fig2}.
On the other hand, this 
resonance pole could be
fictitious since the fit 2$N^*$-3p (dotted curve) with only three poles
are equally acceptable within the fluctuating experimental errors.
Our result suggests that it is important to have more accurate data
in the high $W$ region for a high precision resonance extraction.

\subsection{1$N_0$1$N^*$-3p}
\label{sec3d}

\begin{figure}[t]
\begin{minipage}[t]{75mm}
   \includegraphics[width=75mm]{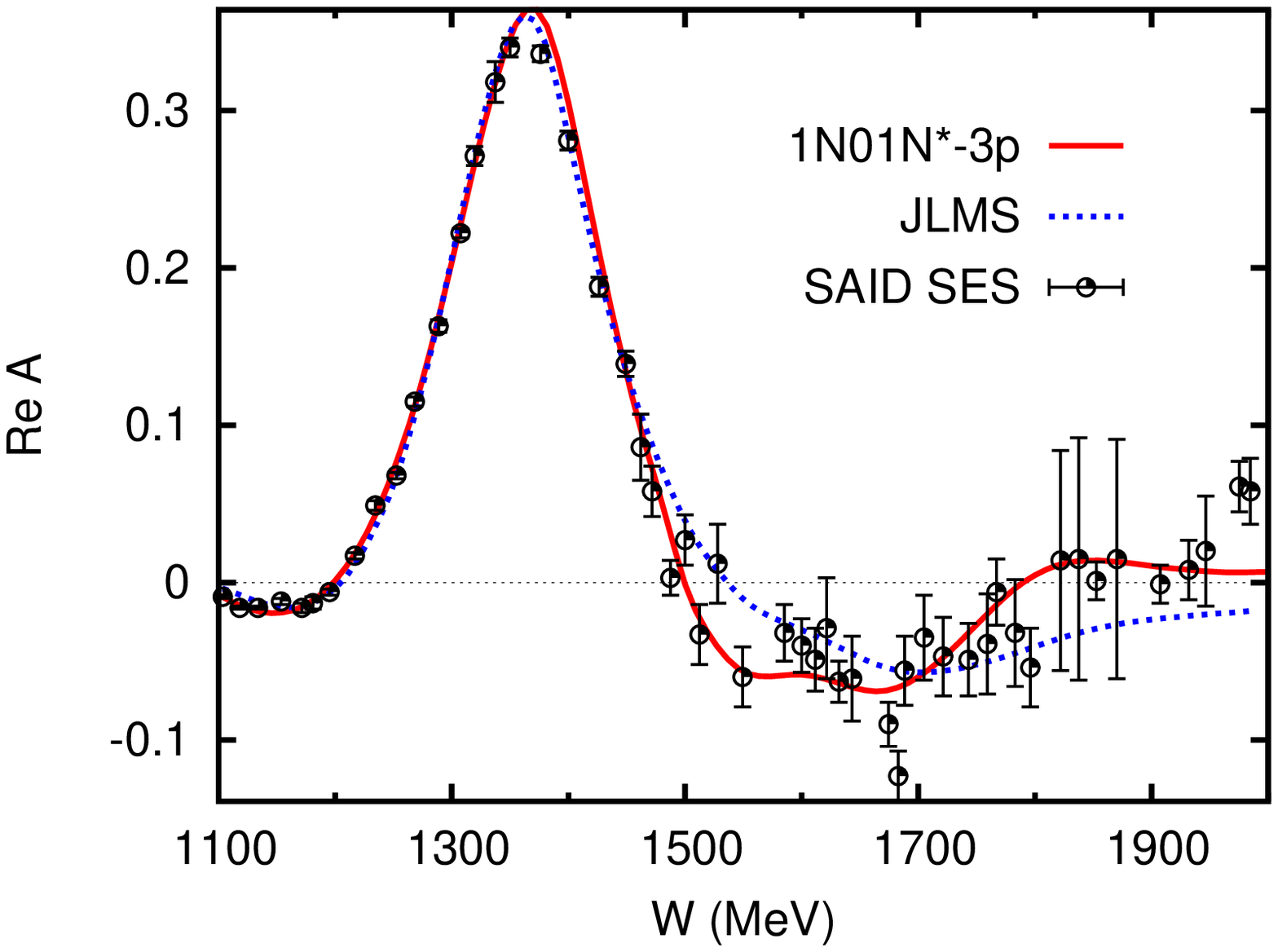}
 \end{minipage}
\begin{minipage}[t]{75mm}
   \includegraphics[width=75mm]{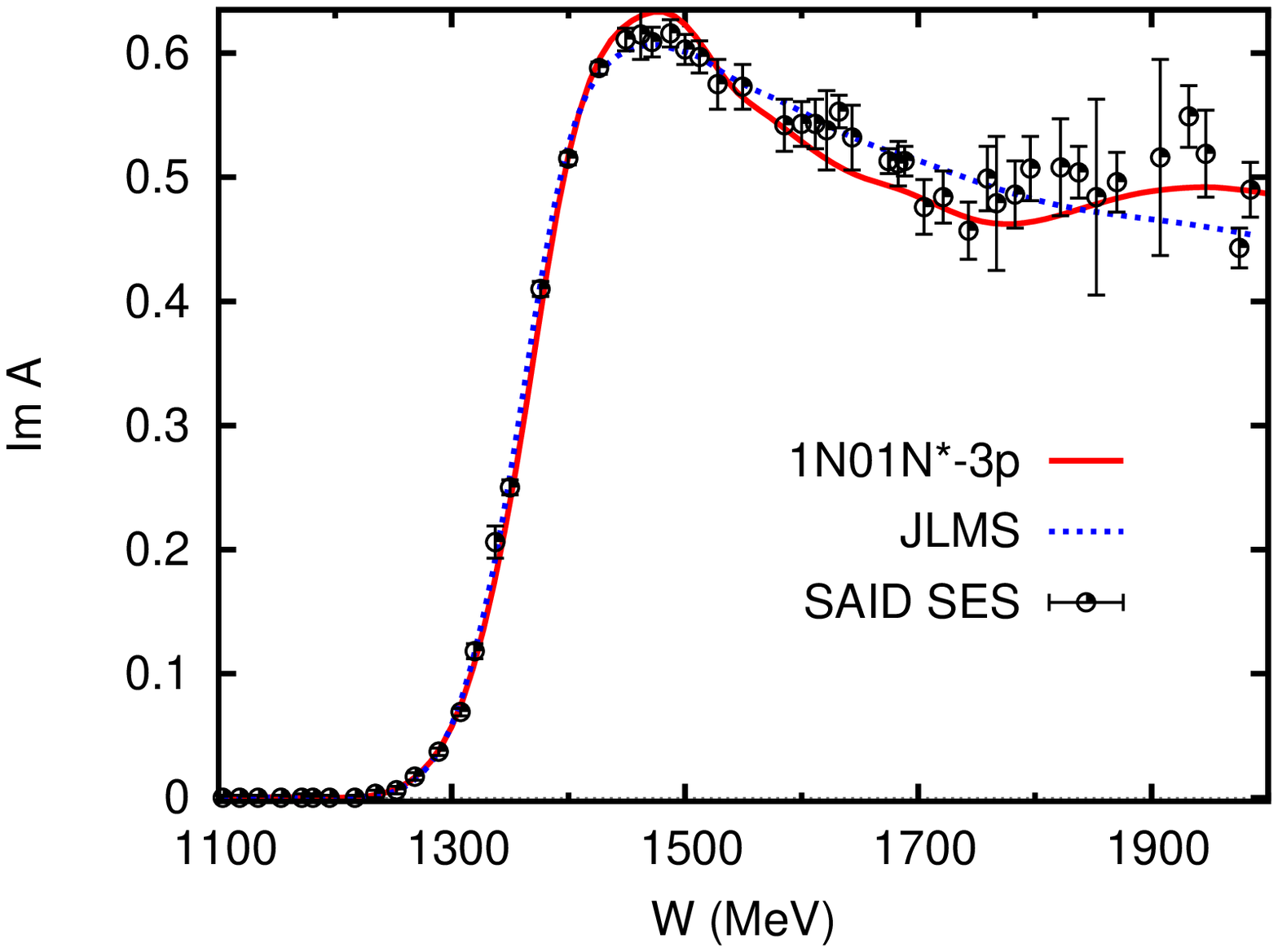}
 \end{minipage}
\caption{\label{fig:p11-fig4}
The real (left panel) and imaginary (right panel) parts of the  $P_{11}$
amplitudes.
}
\end{figure}

Here we show our results obtained with the bare nucleon model, and then
address the question whether difference in the analytic structure of the
$\pi N$ amplitude below $\pi N$ threshold strongly affects the resonance
extractions. 
The bare nucleon model is fitted to SAID-SES, and at the same time, to
the nucleon pole conditions Eqs.~\ref{eq:pole-m}.
Meanwhile, the original EBAC-DCC model has different singular structure
below the $\pi N$ threshold.
The question is whether such differences 
can lead to very different resonance poles.
Our fit of the bare nucleon model is shown in Fig.~\ref{fig:p11-fig4}
and compared with SAID-SES and JLMS.
We see that the two fits agree very well below $W=1.5$ GeV, while
their differences are significant in the high $W$ region.
The corresponding resonance poles are given in Table~\ref{tab:p11-tab1}.
We also see here that the first two
poles near the $\pi\Delta$ threshold are close to those of JLMS.
Our results seem to indicate that these two poles are rather insensitive to
the analytic structure of the amplitude in the region below $\pi N$ threshold, 
and are mainly determined by the data in the region 
$ m_N+m_\pi\leq W \leq 1.6 $ GeV.

\section{Conclusion}

We have examined the stability of the two-pole structure of the Roper
resonance. 
We showed that two resonance poles near the $\pi \Delta$ threshold
are stable against large variations of parameters of  meson-exchange mechanisms
 within EBAC-DCC model~\cite{msl07}.  
This two-pole structure is also obtained 
in an analysis based on a model with the bare nucleon state.
Our results indicate that
the extraction of $P_{11}$ resonances is insensitive to the analytic structure
of the amplitude in the region below $\pi N$ threshold.
Although we did not show a result,
we have also fitted to the old CMB amplitude, which is rather different
from SAID-SES for $W \ge 1.5$~GeV, 
and still found that the Roper two poles are stable.

\begin{theacknowledgments}
This work is supported 
by the U.S. Department of Energy, Office of Nuclear Physics Division, under
Contract No. DE-AC02-06CH11357, and Contract No. DE-AC05-06OR23177
under which Jefferson Science Associates operates Jefferson Lab, and by
the Japan Society for the Promotion of Science,
Grant-in-Aid for Scientific Research(C) 20540270. 
This research used resources of the National Energy Research Scientific Computing Center, which is supported by the Office of Science of the U.S. Department of Energy under Contract No. DE-AC02-05CH11231. 
\end{theacknowledgments}

\bibliographystyle{aipproc}   

\end{document}